\documentclass[11pt]{article}
\usepackage{amsmath,amssymb,color,graphics,epsfig}
\usepackage{graphicx}

\usepackage{amsfonts}

\newcommand{\be}{\begin{equation}}
\newcommand{\ee}{\end{equation}}
\newcommand{\bea}{\setlength\arraycolsep{2pt} \begin{eqnarray}}
\newcommand{\eea}{\end{eqnarray}}
\newcommand{\nn}{\nonumber}
\newcommand{\mc}{\mathcal}
\newcommand{\mm}{\mathrm}

\def\fft#1#2{{\frac{#1}{#2}}}

\def\0{{\sst{(0)}}}
\def\1{{\sst{(1)}}}
\def\2{{\sst{(2)}}}
\def\3{{\sst{(3)}}}
\def\4{{\sst{(4)}}}
\def\5{{\sst{(5)}}}
\def\6{{\sst{(6)}}}
\def\7{{\sst{(7)}}}
\def\8{{\sst{(8)}}}
\def\sst#1{{\scriptscriptstyle #1}}

\begin{document}

\begin{flushright}
\end{flushright}

\vspace{25pt}
\begin{center}
{\large {\bf  Generalised Krylov complexity}}

\vspace{10pt}
 Zhong-Ying Fan$^{1\dagger}$\\

\vspace{10pt}
$^{1\dagger}${ Department of Astrophysics, School of Physics and Material Science, \\
 Guangzhou University, Guangzhou 510006, China }\\


\vspace{40pt}

\underline{ABSTRACT}
\end{center}

The growth rate of Krylov complexity (K-complexity) obeys an upper bound owing to uncertainty relation. However, in this work, by studying generalised notions of K-complexity, we show that
 for slow scramblers, the growth rate of K-complexity at long times cloud be tighter bounded by the generalised counterpart. On the contrary, for fast scramblers, the K-complexity constrains the growth of generalised complexity.

\vfill {\footnotesize  Email: $^\dagger$fanzhy@gzhu.edu.cn\,,}

\thispagestyle{empty}

\pagebreak

\tableofcontents
\addtocontents{toc}{\protect\setcounter{tocdepth}{2}}




\section{Introduction}

  The Krylov complexity ( or K-complexity) is first introduced in \cite{Parker:2018yvk} to describe the Heisenberg evolution of operators $\mc{O}(t)=e^{iHt}\mc{O} e^{-iHt}$, where $\mc{O}$ is an initial operator and $H$ is the lattice Hamiltonian. It is well defined for any quantum system and serves as a new tool to characterize chaotic dynamics. Because of its close relation to dynamical evolution of a many body system, it has attracted a lot of attentions in literature \cite{Caputa:2021sib,Jian:2020qpp,Rabinovici:2020ryf,Dymarsky:2021bjq,Bhattacharjee:2022ave,Bhattacharjee:2022vlt,Afrasiar:2022efk,Pal:2023yik,Muck:2022xfc,Adhikari:2022whf,Adhikari:2022oxr,Fan:2022xaa,Avdoshkin:2022xuw,Camargo:2022rnt,
Hashimoto:2023swv,Iizuka:2023pov,Erdmenger:2023shk,Camargo:2023eev}. The K-complexity can be defined equally well for time evolving quantum states \cite{Balasubramanian:2022tpr,Kundu:2023hbk,Rabinovici:2023yex}. Generalization to open quantum systems is studied in \cite{Liu:2022god,Bhattacharya:2022gbz,Bhattacharjee:2022lzy,Bhattacharya:2023zqt}.

  It was proved in \cite{Hornedal:2022pkc} that the growth rate of K-complexity obeys a dispersion bound $|\partial_t K|\leq 2b_1 \Delta K$, where $b_1$ stands for the first Lanczos coefficient and $\Delta K$ is the variance of K-complexity. In \cite{Fan:2022mdw}, similar bound is established for Krylov entropy (K-entropy) \cite{Barbon:2019wsy} and generalisations of K-complexity. In particular, using a universal logarithmic relation between K-complexity and K-entropy \cite{Fan:2022xaa}, it was show in \cite{Fan:2022mdw} that for irreversible process the growth rate of K-entropy could be tighter bounded by K-complexity. Inspired by this, in this paper, we would like to study generalised K-complexity and demonstrate its relation to the growth of K-complexity.

The paper is organized as follows. In section 2, we briefly review the recursion method and the Lanczos algorithm. In section 3, we examine universal features of generalised complexity at both initial and long times. In section 4, we study the upper bound on the growth rate of generalised complexity. We briefly conclude in section 5.

\section{Brief review of the recursion method}

For a given operator $\mc{O}$, it evolves in time according to
\be \mc{O}(t)=e^{iHt}\mc{O} e^{-iHt}=\sum_{n=0}\fft{(it)^n}{n!}\tilde{\mc{O}}_n\,, \ee
where $\tilde{\mc{O}}_n$ stands for the nested commutators: $\tilde{\mc{O}}_1=[H,\mc{O}_0]\,,\tilde{\mc{O}}_2=[H,\tilde{\mc{O}}_1]\,,\cdots\,,\tilde{\mc{O}}_n=[H,\tilde{\mc{O}}_{n-1}]\,.$
However, it is instrumental to take the operator as a wave function, which evolves under the {\it Liouvillian} $\mc{L}\equiv [H\,,\cdot]$. One has
\be\label{obasis} |\mc{O}(t))=e^{i\mc{L}t}|\mc{O}_0 )= \sum_{n=0}\fft{(it)^n}{n!}|\tilde{\mc{O}}_n )\,, \ee
where $|\tilde{\mc{O}}_n )=\mc{L}^n|\mc{O}_0 ).$ The physical information about operator growth is essentially encoded in the auto-correlation function ( here we have chosen the Wightman inner product $(A|B)\propto\mm{Tr}(A^\dag B)$)
\be C(t):= ( \mc{O}|\mc{O}(t) )=( \mc{O}| e^{i\mc{L}t}  |\mc{O} )\,, \ee
The same information can be equivalently extracted from the moments $\mu_{2n}$
\be
\mu_{2n}:= \langle \mc{O}| \mc{L}^{2n}  |\mc{O}\rangle=\fft{d^{2n}}{dt^{2n}}C(-it)\Big|_{t=0}\,,
\ee
the relaxation function $\phi_0(z)$ and the spectral density $\Phi(\omega)$
\bea
&& \phi_0(z):= \int_{0}^{+\infty}dt\, e^{-z t} C(t)\,,\nn\\
&& \Phi(\omega):= \int_{-\infty}^{+\infty}dt\, e^{-i\omega t} C(t) \,.
\eea
This provides four equivalent ways (which are linearly related) to describe operator growth. It turns out that the Lanczos algorithm provides a fifth equivalent description. In general, the original basis $\{|\tilde{\mc{O}}_n )\}$ is not orthogonal. One can construct an orthonormal basis using the Gram-Schmidt scheme, as in ordinary quantum mechanics. Starting with a normalized vector $|\mc{O} )$, the first vector is given by $|\mc{O}_1):=b_1^{-1}\mc{L}|\mc{O})$, where $b_1:=(\mc{O}\mc{L}|\mc{L}\mc{O} )^{1/2}$. For the $n-$th vector, one has inductively
\bea\label{kbasis}
&&|A_n):=\mc{L}|\mc{O}_{n-1})-b_{n-1}|\mc{O}_{n-2})\,,\nn\\
&&|\mc{O}_n ):=b_n^{-1}|A_n )\,,\quad b_n:=( A_n|A_n )^{1/2}\,.
\eea
If at the $k$-th step, $b_k=0$, then the recursion stops. The output of this procedure is a set of orthomornal basis $\{|\mc{O}_n )\}$, referred to as {\it Krylov basis} and a sequence of positive numbers $\{b_n\}$, referred to as {\it Lanczos coefficients}. Note that the coefficients have units of energy and can be used to measure time in the Heisenberg evolution. It should be emphasized that while the Lanczos coefficients $\{ b_n \}$ contains equivalent information about the operator growth, it is nonlinearly related to the auto-correlation function as well as the moments, the relaxation function and the spectral density.

In Krylov basis, evolution of the operator wave function $|\mc{O}(t))$ can be formally written as
\be\label{newbasis} |\mc{O}(t)):=\sum_{n=0}i^n\varphi_n(t)|\mc{O}_n)\,,  \ee
where $\varphi_n(t)$ stands for a discrete set of (real) wave functions and $p_n\equiv \varphi_n^2$ can be interpreted as probabilities. One has the normalization $\sum_{n=0}^{\infty}\varphi^2_n(t)=1 $.
The Heisenberg evolution of $\mc{O}(t)$ gives rise to a discrete set of equations
\be\label{varphi} \partial_t\varphi_n=b_n\varphi_{n-1}-b_{n+1}\varphi_{n+1} \,,\ee
subject to the boundary condition $\varphi_n(0)=\delta_{n0}$ and $b_0=0=\varphi_{-1}(t)$ by convention. Notably the auto-correlation function is simply given by $C(t)=\varphi_0(t)$.

The Krylov complexity (K-complexity) and the Krylov entropy (K-entropy) are defined respectively as
\bea  K&=& \sum_{n=0}n\, \varphi_n^2 \,,\\\nn
     S_K&=&\sum_{n=0}-\varphi_n^2\,\mm{ln}\varphi_n^2 \,,\eea
In this paper, we are interested in a set of generalisation of K-complexity: the complexity with degree $\delta$
\be  K_\delta=\sum_{n=0}n^\delta \varphi_n^2 \,, \ee
where $\delta\geq 1$ is a positive integer. These quantities have not been carefully studied in literature. It was shown in \cite{Fan:2022mdw} that all these quantities obey a dispersion bound
\be |\partial_t A|\leq 2b_1\Delta A \,,\label{dispersion}\ee
where $A$ collectively denotes the Krylov quantities and $\Delta A$ stands for the variance. This is a quantum ultimate limit for all Krylov quantities (defined properly) in Heisenberg evolution.

However, in this work, we will show that for slow scramblers the generalised Krylov complexities can actually constrain the growth rate of K-complexity more stringently than the dispersion bound at late times (to avoid confusion, we refer the system to as fast scrambler if the Lanczos coefficient $b_n$ grows asymptotically linearly and if not, we refer it to as slow scrambler). The result, together with universal features of all complexities may help to comprehend holographic dual of K-complexity in the near future.

\section{Universal features of generalised complexity}

\subsection{Initial growth}
Let us first study initial behavior of the generalised K-complexity. From the discrete Schr\"{o}dinger equation (\ref{varphi}), together with the boundary condition, we find that at initial times, all the complexity turns out to be a series of even powers of time, namely
\be K_\delta=\sum_{k=1}^\infty c_{2k}^{(\delta)}\, t^{2k} \,.\ee
This should the case because: $1)$ the lattice Hamiltonian $H$ is Hermitian; $2)$ the operator $\mc{O}$ under consideration is Hermitian; $3)$ the inner product for the Krylov space is chosen as $(A|B)\propto\mm{Tr}(A^\dag B)$. Here there are two interesting features. First, to leading order all the complexity shares exactly the same behavior. We find
\be c_2^{(\delta)}=\mu_2\quad \Longrightarrow \quad K_\delta=\mu_2t^2+\cdots \,,\ee
where $\mu_2=b_1^2$ is the first moment. However, inclusion of the subleading order terms, different complexity will generally behave different. For example, for K-complexity
\be c_4=\fft{b_1^2}{6}\big( b_2^2-2b_1^2 \big)\,,\quad c_6=\fft{b_1^2}{180}\big( 3b_2^2b_3^2-7b_2^4+b_1^2b_2^2+8b_1^4\big)  \,,\ee
whereas for $K_2$
\be c_4^{(2)}=\fft{b_1^2}{3}\big( 2b_2^2-b_1^2 \big)\,,\quad c_6^{(2)}=\fft{b_1^2}{90}\big( 9b_2^2b_3^2-11b_2^4-7b_1^2b_2^2+4b_1^4\big) \,.\ee
Generally speaking, for any degree complexity, the coefficient at the $2k$-th order $c_{2k}^{(\delta)}$ is given by the Lanczos coefficients $(b_1^2\,,b_2^2\,,\cdots\,,b_k^2)$, namely
\be c_{2k}^{(\delta)}=c_{2k}^{(\delta)}(b_1^2\,,b_2^2\,,\cdots\,,b_k^2)\,.\ee
In the other way around, provided initial behavior of any complexity, all Lanczos coefficients can be solved order by order from the series coefficients. In other words, any generalised K-complexity provides an alternate (and equivalent) description for operator growth\footnote{However, for a general choice of inner product, there will be two sets of Lanczos coefficients to describe the operator growth. In this case, one needs two different complexities at least to extract the Lanczos coefficients. This can be viewed another reason to study generalised K-complexity.}. This partly explains why these quantities deserve careful studies.

\begin{figure}
  \centering
  \includegraphics[width=250pt]{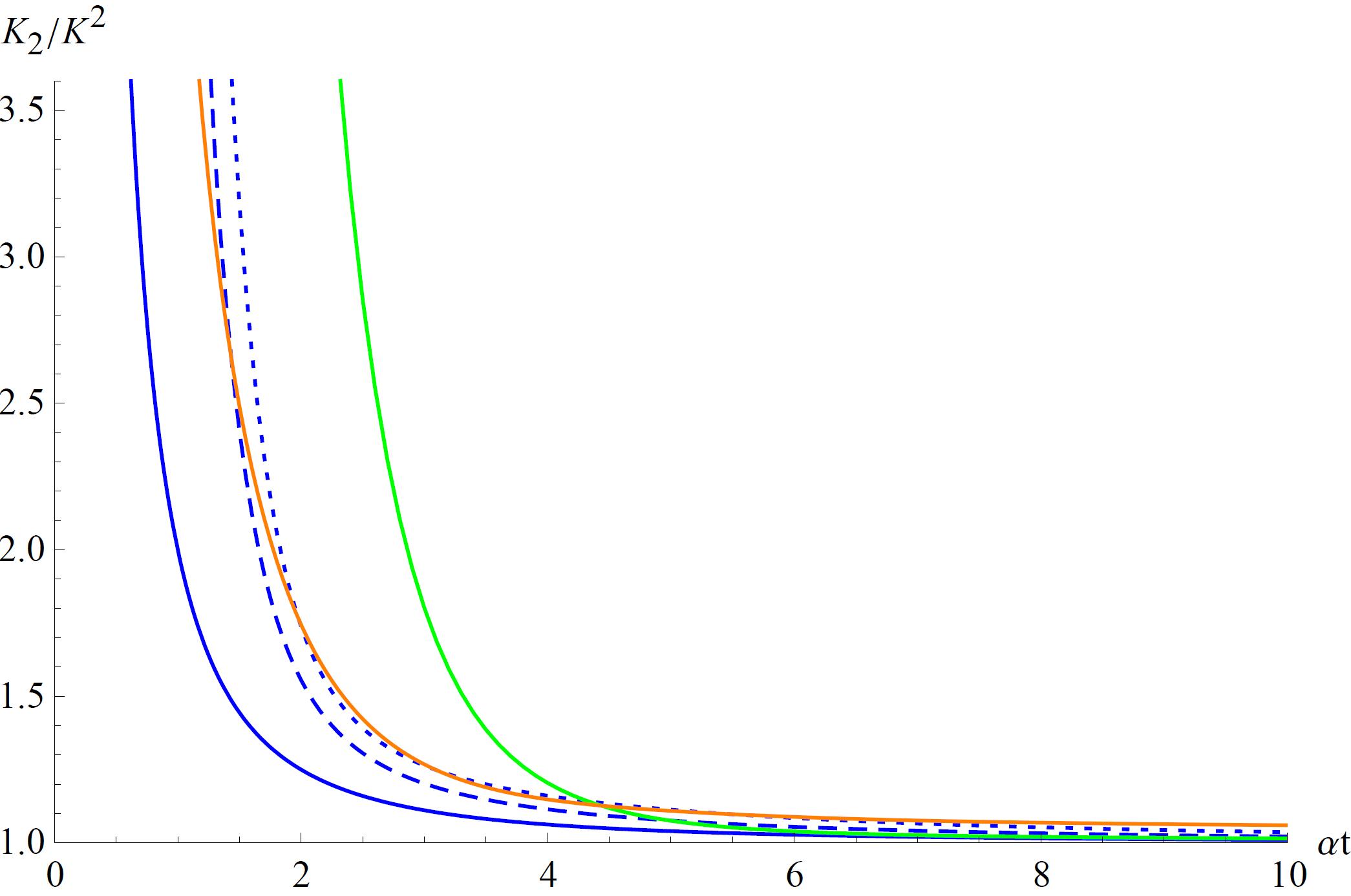}
  \caption{The ratio of complexities $K_2/K^2$ for various models. For integrable models $b_n=\alpha n^\gamma$, $\gamma=1/2$ (blue), $\gamma=2/3$ (dashed), $\gamma=3/4$ (dotted). The green line characterizes $1d$ chaotic model $b_n=\alpha n/\mm{ln}(n+1)$ whereas the orange line describes a constant Lanczos coefficient $b_n=\alpha$. In all these cases, the ratio $K_2/K^2$ approaches to unity at late times. The result is valid to generalised complexity with higher degrees.}
  \label{k2}
\end{figure}

\subsection{The long time limit}
For our purpose, we are more interested in examining the behavior of complexity at long times. Qualitatively, this can be analyzed using continuum limit approximation (details are presented in the Appendix A). It turns out that  at long times the wave functions move ballistically with a velocity $v=2b_n$ and the K-complexity measures the location of the wave function $K\sim n$ approximately. The result can be written as
\be \fft{dK}{dt}\Big|_{t\rightarrow \infty}=2b(K) \,.\label{continuum1}\ee
Thus, given the asymptotic growth of the Lanczos coefficient, one can extract the behavior of K-complexity at late times and vice versa. A known example is if the Lanczos coefficient grows asymptotically linearly $b_n=\alpha n$, the K-complexity will increase exponentially $K\sim e^{2\alpha t}$. The logic can be reversed without any difficulty from (\ref{continuum1}). This applies to general cases since (\ref{continuum1}) is only a first order equation.

For generalised complexity, the continuum limit implies at long times $K_\delta \sim K^\delta$, where the coefficient is undetermined by coarse grained analysis. However, from numerical results we find that for any slow scrambler, the coefficient approaches to unity, as shown in Fig. \ref{k2}. However, for fast scramblers, this is no longer the case. For example, for SYK like model which has $b_n=\alpha\sqrt{n(n-1+\xi)}$, we find in the long time limit
\be K_\delta=\fft{(\xi)_\delta}{\xi^\delta} K^\delta \,,\label{genefast}\ee
where $(\xi)_\delta=\xi(\xi+1)\cdots(\xi+\delta-1)$ is the Pochhammer symbol. Clearly the proportional coefficient is generally not equal to unity. We refer the readers to Appendix B for more details about the complexity of this model. 

 It turns out that the above results for both fast scramblers and slow scramblers can be partly explained from known features of K-complexity. As an example, we focus on $K_2$. By definition $K_2=K^2+(\Delta K)^2$, where $\Delta K$ stands for the variance of K-complexity. It is shown \cite{Hornedal:2022pkc} that for fast scramblers, the dispersion bound of K-complexity is saturated asymptotically. This gives when $t\rightarrow \infty$
\be \Delta K=\partial_t K/2b_1=K/\hat{b}_1\quad \quad \Longrightarrow\quad  K_2=\big(1+\fft{1}{\hat{b}_1^2} \big)K^2\,,\label{k2fast}\ee
where $\hat{b}_1=b_1/\alpha$. This is universal to fast scramblers. For instance, for SYK-like model $\hat{b}_1=\sqrt{\xi}$ so that (\ref{genefast}) reduces to (\ref{k2fast}). On the other hand, for slow scramblers, the continuum limit implies the complexity still grows fast at long times:  the variance $\Delta K$ is in the same order of $\partial_t K/b_1$ but  $\Delta K/K\sim \partial_t K/b_1K\sim 1/t^\sigma$ decays at late times, where $\sigma$ is some positive constant. This illustrates that in the long time limit, the contribution of the relative variance $\Delta K/K$ could be safely neglected and hence $K_2=K^2$. This explains the results depicted in Fig. \ref{k2}. The same argument can be extended to higher degree complexities. We may conclude that asymptotic behavior of the ratio $K_\delta/K^\delta$ at long time limit serves as a new tool to distinguish between slow scramblers and the fast ones.

\section{The upper bound on complexity growth}

\subsection{Relation to Krylov entropy}\label{enkk}

It was first established in \cite{Fan:2022xaa} that for semi-infinite chains, there exists a logarithmic relation between K-complexity and K-entropy at long times
\be S_K=\eta\, \mm{ln}K+\cdots \,,\label{logk}\ee
where $\eta$ is a constant depending on models under considerations. In particular, if the Lanczos coefficient asymptotically grows as $b_n=\alpha n^\gamma\,,1/2\leq \gamma\leq 1$, then $\eta=\gamma$. Besides, the growth rate of K-entropy also obeys a dispersion bound $ |\partial_t S_K|\leq 2b_1\Delta S_K$. However, the bound turns out to be too loose since $\partial_t S_K$ generally decays in a power law while the variance $\Delta S_K$ approaches to a constant \cite{Fan:2022mdw}
. Interestingly, based on the logarithmic relation $(\ref{logk})$ and the dispersion bound of K-complexity, it was shown \cite{Fan:2022mdw} that the K-complexity provides a tighter bound on the growth rate of K-entropy
\be t\rightarrow \infty\,,\quad |\partial_t S_K|\leq 2b_1\,\eta \Delta K/K\leq 2b_1\Delta S_K \,.\label{boundsk}\ee
Despite that the result holds in the long time limit, it characterizes the late time behavior of $S_K$ very well for physically interesting cases. This inspires us to study whether the growth rate of K-complexity can be bounded more stringently as well according to its relation to generalised complexity. We find the answer is ``yes".

\begin{figure}
  \centering
  \includegraphics[width=250pt]{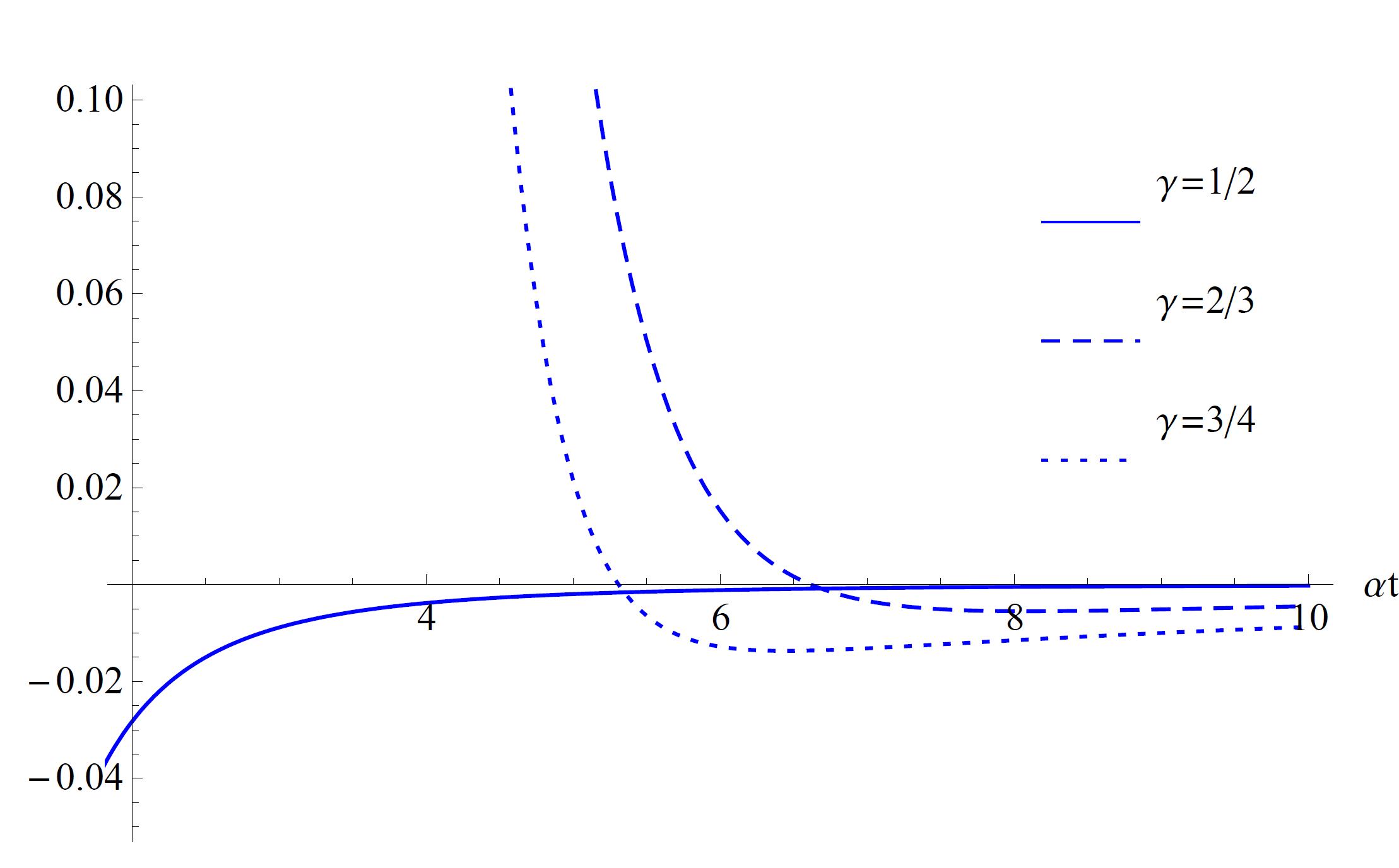}
  \caption{The plot of the difference $(\Delta K/K-\Delta K_2/2K_2)$ for integrable models $b_n=\alpha n^\gamma$. The bound (\ref{master}) is saturated asymptotically.}
  \label{diff}
\end{figure}

Now the logarithmic relation (\ref{logk}) can be extended to
\be  S_K=\eta\, \mm{ln}K+\cdots=\eta_\delta\,\mm{ln}K_\delta+\cdots \,,\label{logkd}\ee
where $\eta_\delta=\eta/\delta$. This implies to leading order
\be \partial_t K/K=\partial_t K_\delta/\delta K_\delta \,.\ee
However, for fast scramblers, the dispersion bound is saturated asymptotically $\partial_t K=2b_1\Delta K$. This implies
\be t\rightarrow \infty\,,\quad \Delta K/K\leq \Delta K_\delta/\delta K_\delta \,,\label{master}\ee
where the inequality follows from the dispersion bound $\partial_t K_\delta \leq 2b_1\Delta K_\delta$. While the relation (\ref{master}) is derived for fast scramblers, we will show that the bound is saturated by slow scramblers.

To gain an intuition about the result (\ref{master}), let us first study the SYK-like model explicitly. Given the relation (\ref{genefast}), we deduce
\bea
\Big( \fft{\Delta K_\delta }{K_\delta } \Big)^2&=&\fft{K_{2\delta}-K_\delta^2}{K_\delta^2}=\fft{(\xi)_{2\delta}}{\big[ (\xi)_\delta \big]^2 }-1 \nn\\
&=&\fft{(\delta+\xi)(\delta+\xi+1)\cdots (\delta+\xi+\delta-1)}{\xi(\xi+1)\cdots(\xi+\delta-1)}-1\nn\\
&\geq & \fft{\delta \sum_{k=0}^{\delta-1}\fft{\delta (\xi)_\delta}{\xi+k}}{(\xi)_\delta}\geq \fft{\delta^2}{\xi}\,.
\eea
In addition, according to (\ref{k2fast})
\be \Big( \fft{\Delta K}{K} \Big)^2=\fft{1}{\hat{b}_1^2}=\fft{1}{\xi} \,.\ee
Combining the above results, one indeed arrives at (\ref{master}). This gives us confidence that the result is reasonable.
\begin{figure}
  \centering
  \includegraphics[width=210pt]{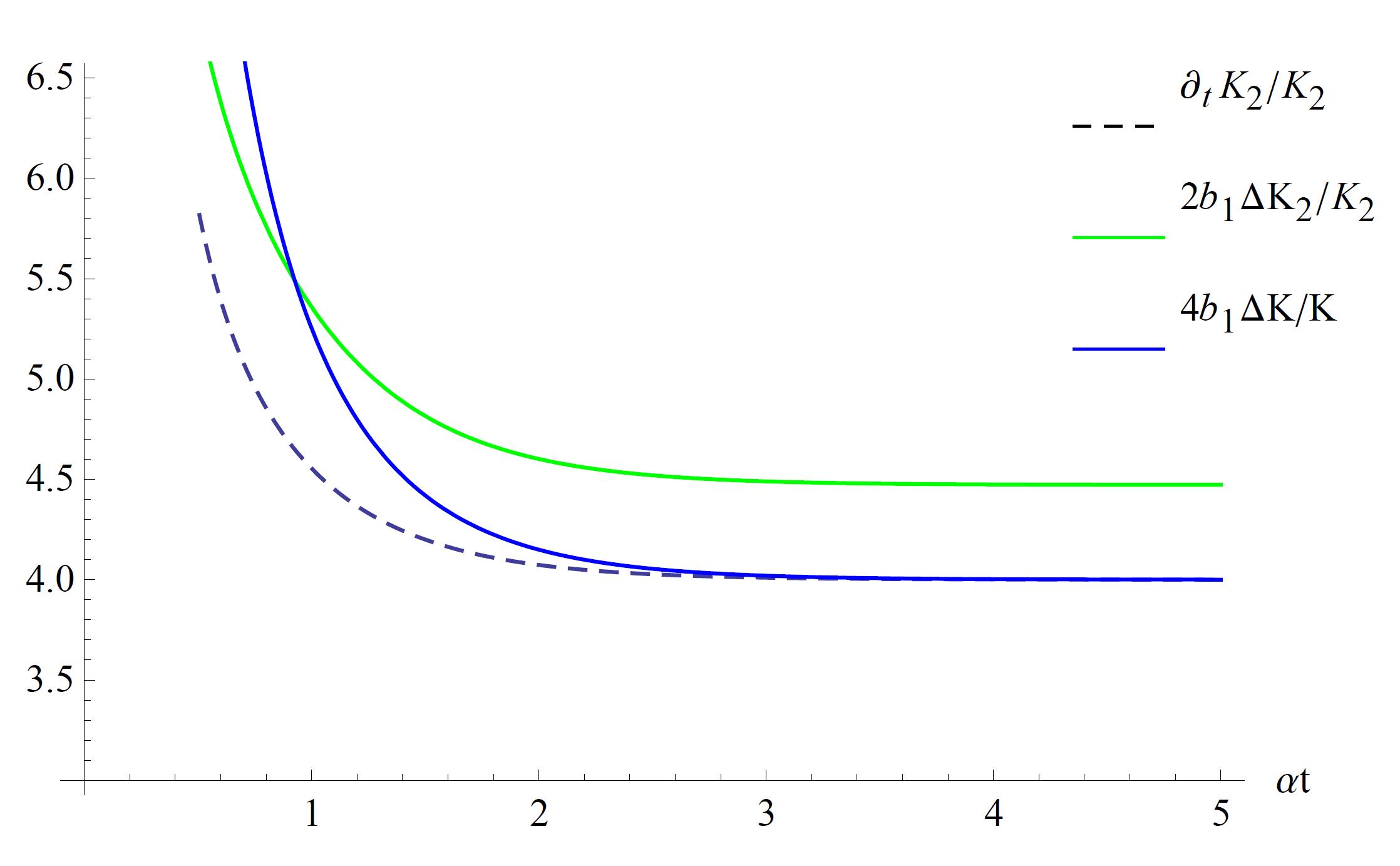}
  \includegraphics[width=210pt]{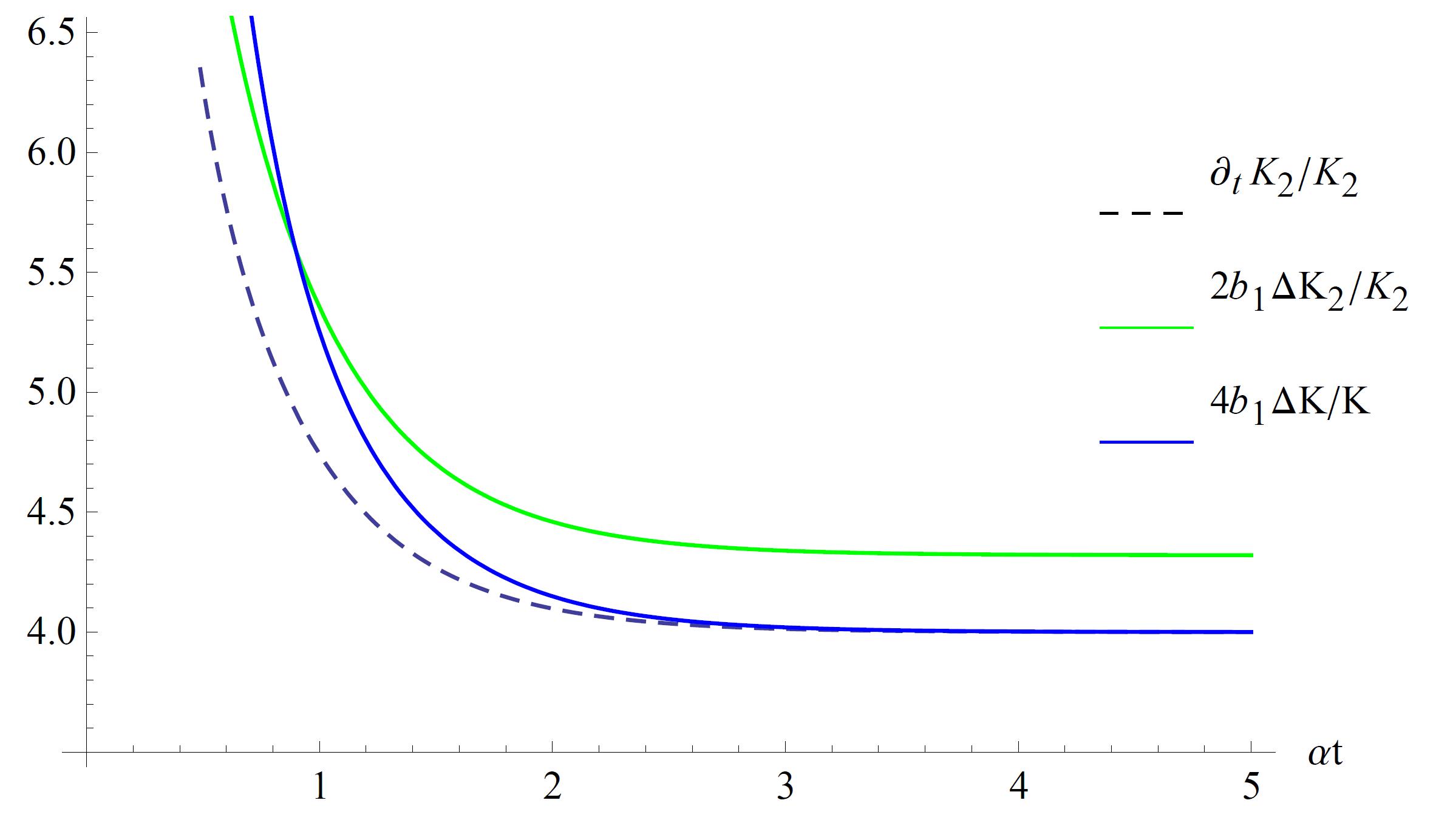}
  \caption{The growth rate of the generalised complexity $ K_2$ for SYK-like model. $\xi=1$ (left panel) and $\xi=2$ (right panel). Compared to the dispersion bound, the growth rate is tighter bounded by the variance of K-complexity, according to (\ref{boundkd}).}
  \label{syk}
\end{figure}
On the other hand, for slow scramblers, the continuum limit analysis implies that the relative variance of generalised complexity generally decays in the long time limit $\Delta K_\delta/K_\delta \sim \partial_t K_\delta/b_1K_\delta \sim 1/t^\sigma$
and hence $\Delta K/K\sim \Delta K_\delta/\delta K_\delta\rightarrow 0$. This implies that slow scramblers saturate (\ref{master}) asymptotically. This passes a variety of numerical tests, see Fig. \ref{diff}.

Given the relation (\ref{master}), one may arrive at a tighter bound on the growth rate of generalised complexity
\be \partial_t K_\delta\leq 2\delta b_1 \Delta K K_\delta/K\leq 2b_1\Delta K_\delta \,.\label{boundkd}\ee
However, the result is only valid in the long time limit. Whether it can be extended to finite times depends on details of lattice models. Using numerical results, we show that (\ref{boundkd}) is only valid to fast scramblers whereas for slow scramblers the growth rate of K-complexity is bounded as
\be \partial_t K\leq 2 b_1 \Delta K_\delta K/\delta K_\delta \leq 2b_1\Delta K \,,\label{boundk}\ee
which is the desired result we search.

\subsection{Numerical results}
\begin{figure}
  \centering
  \includegraphics[width=210pt]{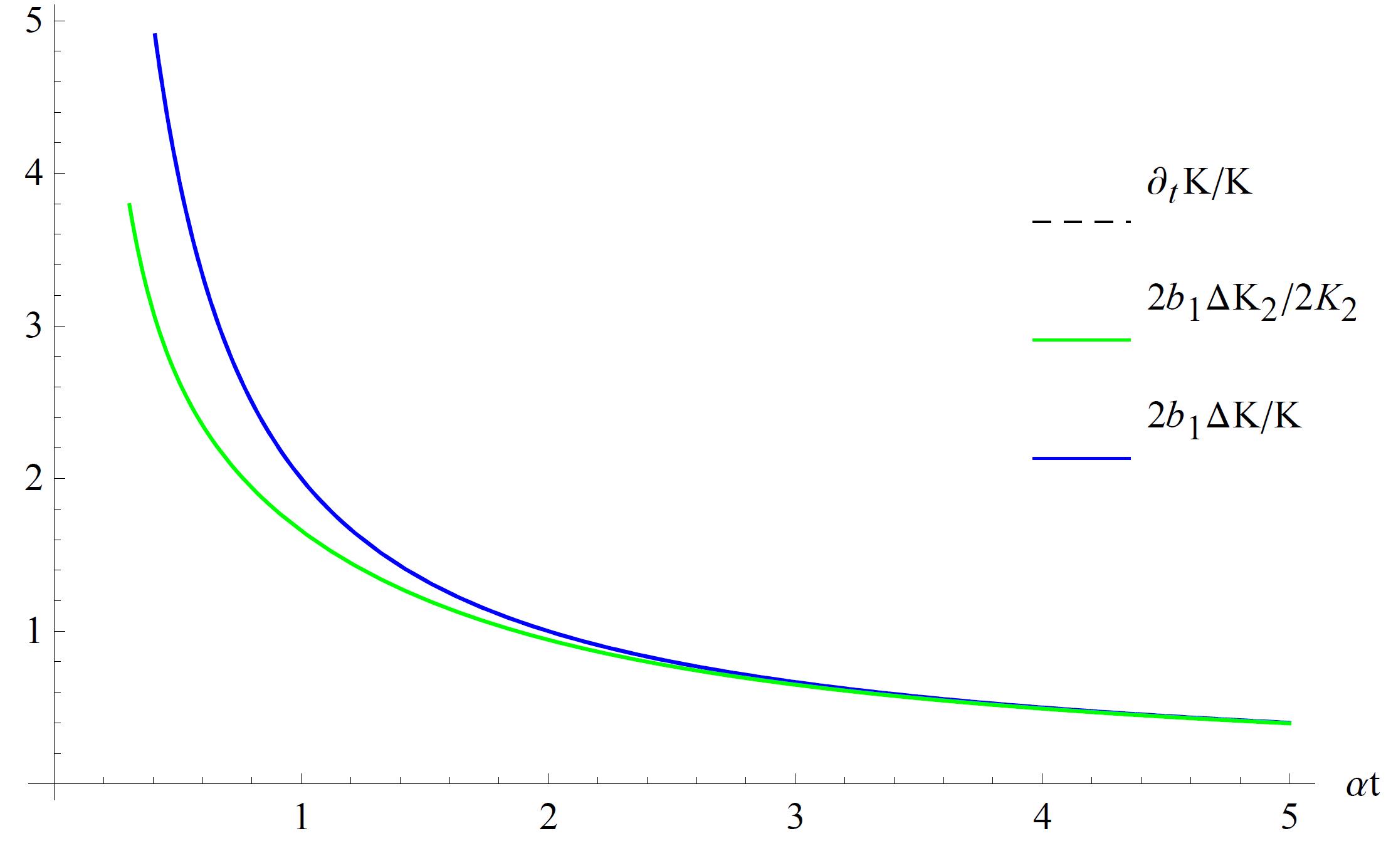}
  \includegraphics[width=210pt]{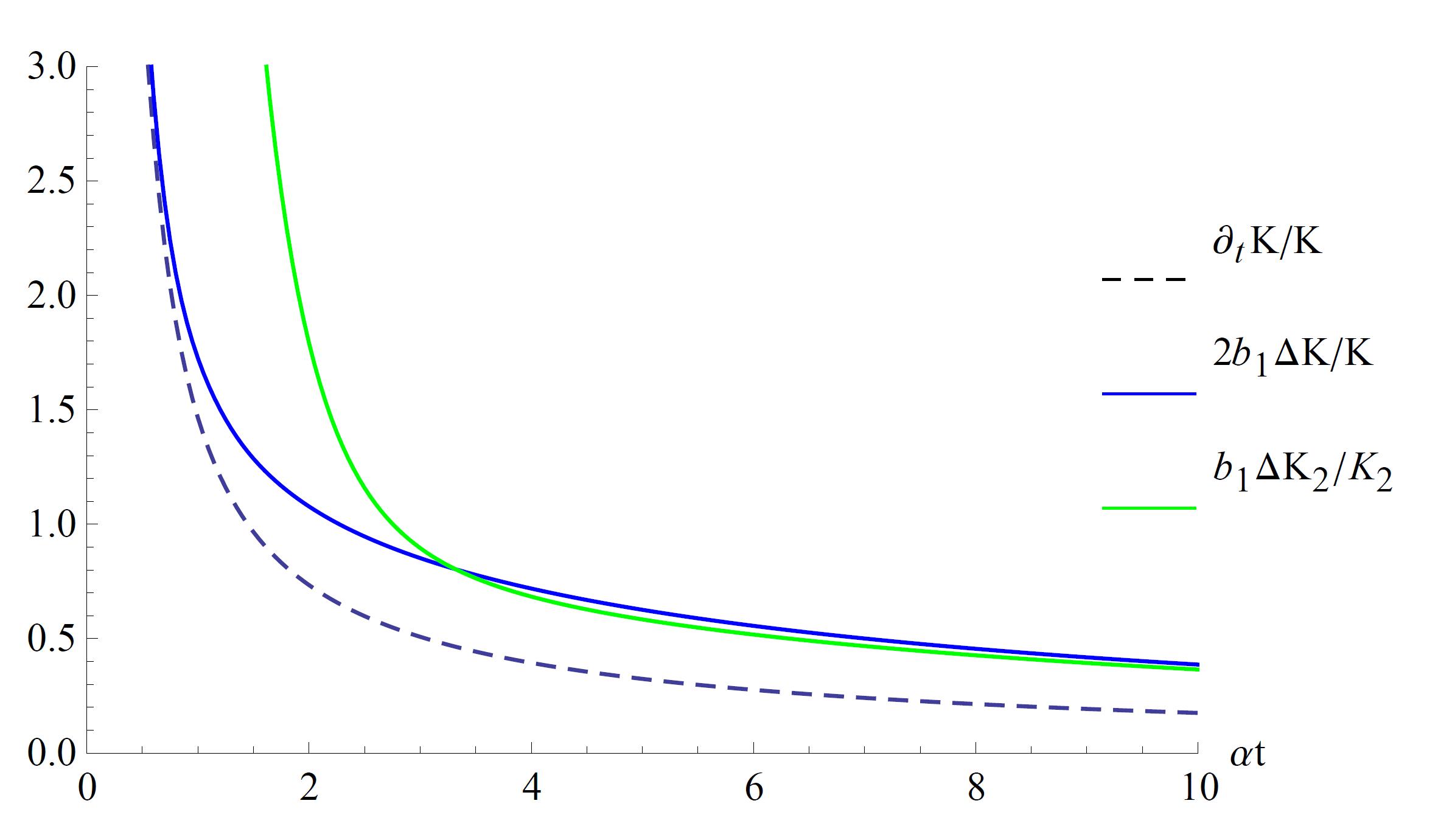}
  \caption{The growth rate of K-complexity for integrable models in which the Lanczos coefficient grows asymptotically as $b_n\rightarrow \alpha\sqrt{n}$. Left panel: the exact Heisenberg-Weyl case. The dispersion bound of K-complexity is saturated in the evolution but is lower bounded by the generalised complexity. Right panel: the Lanczos coefficient is specified as $b_n=\alpha \sqrt{n+5}$. The growth rate of K-complexity is slightly tighter bounded by the variance of the generalised complexity.}
  \label{hw}
\end{figure}

Now let us test the bound (\ref{boundkd}) or (\ref{boundk}) numerically for a number of cases. Without loss of generality, we focus on $K_2$. Similar results can be obtained for higher degree complexity. In Fig. \ref{syk}, we present the growth rate of $K_2$ for SYK-like model: $\xi=1$ (left panel) and $\xi=2$ (right panel). It is clear that the bound (\ref{boundkd}) is satisfied. In particular, the tighter bound given by the K-complexity describes the behavior of $K_2$ very well at long times.

\begin{figure}
  \centering
  \includegraphics[width=210pt]{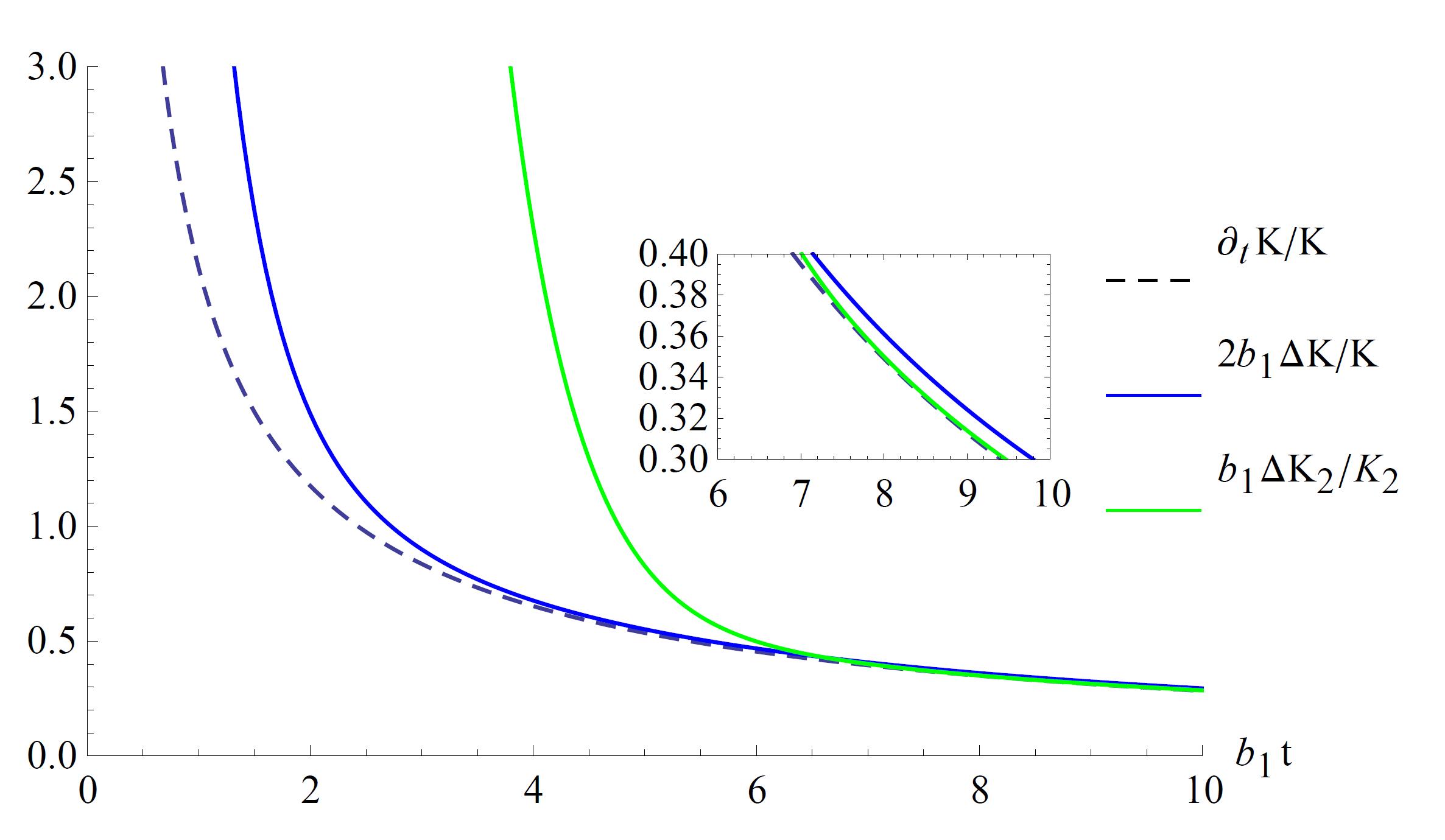}
  \includegraphics[width=210pt]{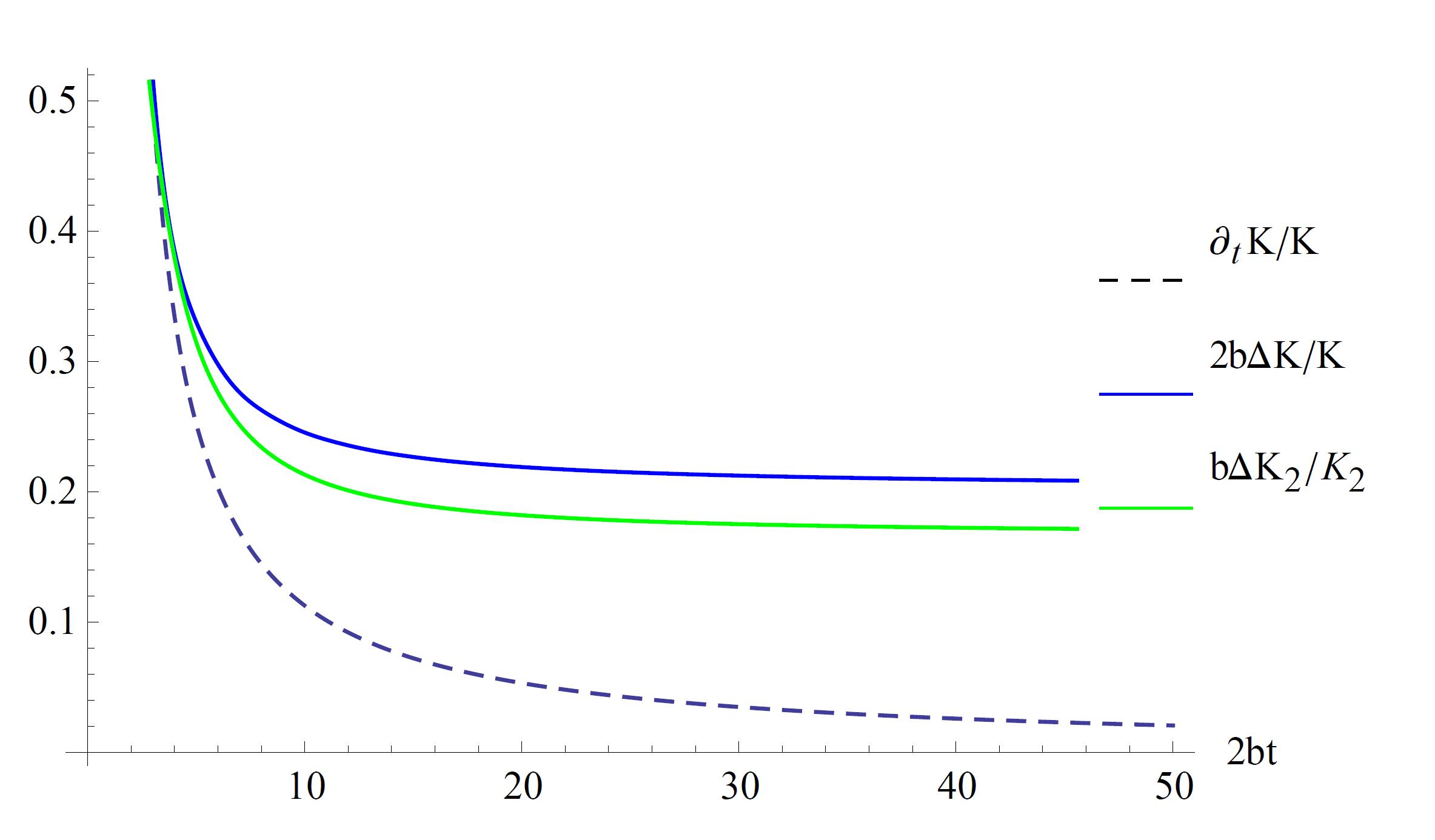}
  \caption{The growth rate of K-complexity for the integrable model $b_n=\alpha n^{2/3}$ (left panel) and that with bounded support $b_n=b$ (right panel). In both cases, the growth rate of K-complexity is tighter bounded by the generalised complexity. }
  \label{more}
\end{figure}

Next, we study integrable models $b_n=\alpha n^\gamma$. It was known that for the Heisenberg-Weyl case $\gamma=1/2$, the dispersion bound on the growth rate of K-complexity is exactly saturated and hence (\ref{boundk}) is in fact incorrect. Instead, we find
\be  2 b_1 \Delta K_\delta K/\delta K_\delta \leq \partial_t K= 2b_1\Delta K \,,\ee
as shown in the left panel of Fig. \ref{hw}. However, in many lattice models, the Lanczos coefficient just behaves asymptotically as $b_n\rightarrow \alpha\sqrt{n}$. In the right panel of Fig. \ref{hw}, we show that in this case the growth rate of K-complexity is slightly tighter bounded by the generalised complexity, satisfying (\ref{boundk}). We test this for more integrable models as well as other slow scramblers, see Fig. \ref{more}. As far as we check, this is always correct. This means that while K-complexity grows fast at long times $\partial_t K\sim 2b_1 K$, its growth rate can in fact be better estimated using the generalised complexity, according to (\ref{boundk}).

\section{Conclusion}

In this paper, we studied a set of generalised Krylov complexity for operator growth. We demonstrate their universal features at both initial times and long times using half-analytical technique as well as numerical results. In particular, by using the logarithmic relation to the Krylov entropy, we establish an inequality (\ref{master}) between the variance of the K-complexity and the generalised notions which holds in the long time limit. Extending the result to finite (but long) times, we show that for fast scramblers, the K-complexity constrains the growth of generalised complexity more stringently than the dispersion bound. However, for slow scramblers, the growth rate of K-complexity is tighter bounded by the generalised complexity in the other way around. Our results enlarge the zoo of Krylov quantities and may shed new light on the future research in this field.

\section*{Acknowledgments}

Z.Y. Fan was supported in part by the National Natural Science Foundations of China with Grant No. 11805041, No. 11873025.

\appendix
\section{Continuum limit analysis at long times}

For semi-infinite chains, continuum limit analysis is good at capturing the leading long time behaviors of Krylov quantities using coarse grained wave functions.
We first introduce a lattice cutoff $\epsilon$ and define a coordinate $x=\epsilon n$ as well as a velocity $v(x)=2\epsilon b_n$. The interpolating wave function is simply defined as $\varphi(x\,,t)=\varphi_n(t)$. The wave equation (\ref{varphi}) now becomes
\be \partial_t\varphi(x\,,t)=\fft{1}{2\epsilon}\Big[ v(x)\varphi(x-\epsilon)-v(x+\epsilon)\varphi(x+\epsilon) \Big] \,. \ee
Expansion in powers of $\epsilon$ yields to leading order
\be\label{expansion} \partial_t\varphi=-v(x)\partial_x\varphi-\fft12 \partial_x v(x)\varphi+O(\epsilon) \,.\ee
This is a chiral wave equation with a position-dependent velocity $v(x)$ and mass $\fft12 \partial_x v(x)$. The equation is much simplified in a new frame $y$ defined as $y=\int dx/v(x)$ and a rescaled wave function
\be \psi(y\,,t)=\sqrt{v(y)}\,\varphi(y\,,t) \,.\ee
One finds
\be\label{chiralwave} (\partial_t+\partial_y)\psi(y\,,t)=0+O(\epsilon) \,.\ee
The general solution is simply given by
\be \psi(y\,,t)=\psi_i(y-t) \,,\ee
where $\psi_i(y)=\psi(y\,,0)$ stands for the initial amplitude. The result implies that to leading order the wave function simply moves ballistically at long times. Note the normalization condition
\be 1=\sum_{n}|\varphi_n(t)|^2=\fft{1}{\epsilon}\int \mm{d}x\, \varphi^2(x\,,t)=\fft{1}{\epsilon}\int \mm{d}y\, \psi^2(y\,,t)  \,.\ee
Evaluation of the complexities yields
\bea\label{cksk}
K(t)&=&\sum_{n}n |\varphi_n(t)|^2=\fft{1}{\epsilon}\int\mm{d}y\, \fft{x(y+t)}{\epsilon}\,\, \psi_i^2(y) \,,\\
K_\delta(t)&=&\sum_{n}n^2 |\varphi_n(t)|^2=\fft{1}{\epsilon}\int\mm{d}y\, \fft{x^\delta(y+t)}{\epsilon^\delta}\,\, \psi_i^2(y) \,.
\eea
Using these results, once the transformation between the two frames is known, we are able to extract the leading time dependence of the quantities immediately. However, for our purpose, we do not need these details. In fact, taking the long time limit, (\ref{cksk}) already tells us $K\sim x(t)/\epsilon$ and $K_\delta\sim x^\delta(t)/\epsilon^\delta=K^\delta$, up to a proportional coefficient.

To proceed, we evaluate the variance of complexities. Using Taylor expansion, we deduce to leading order at long times
\be \Delta K_\delta\simeq \fft{\delta x^{\delta-1}x'}{\epsilon^\delta}\sqrt{Y_2-Y_1^2}\sim \partial_t K_\delta/b_1 \,,\ee
where $Y_n\equiv \fft{1}{\epsilon}\int dy\,y^n\,\psi^2_i(y)$ and we have ignored higher order terms (these terms are in the same order of $x'(t)/\epsilon$ for fast scramblers but this does not influence our discussions). This implies that all complexities grow fast at long times $\partial_t K_\delta \sim b_1 \Delta K_\delta$ even if the bound is not saturated asymptotically. For fast scramblers, this is not a surprise since $K_\delta \sim \Delta K_\delta\sim e^{2\delta \alpha t}$. However, for slow scramblers it implies the relative variance $\Delta K_\delta/K_\delta$ should decay and hence can be neglected at long times.

Using the same analysis, the variance of K-entropy can be evaluated as
\be \Delta S_K=\fft{1}{\epsilon}\int\mm{d}y\, \psi_i^2(y)\,\mm{ln}^2{\psi_i^2(y)}-\fft{1}{\epsilon^2}\Big[\int\mm{d}y\, \psi_i^2(y)\,\mm{ln}{\psi_i^2(y)}\Big]^2 \,,\label{deltask}\ee
which is of order unity. Recall the logarithmic relation $S_K=\eta_\delta \,\mm{ln}K_\delta+\cdots$. Then the dispersion bound of K-entropy implies
\be \eta_\delta \Delta K_\delta/K_\delta\leq \Delta S_K \,.\ee
This generalizes the master result in \cite{Fan:2022mdw} for K-complexity.

\section{Generalised complexity for SYK-like model}

Here we present the results for generalised complexity for SYK-like model which has the Lanczos coefficient $b_n=\omega \sqrt{n(n-1+\xi)}$. The wave function is given by \cite{Parker:2018yvk}
\be \varphi_n(t)=\sqrt{\fft{(\xi)_n}{n!}}\,\fft{\tanh^n(\omega t)}{\cosh^{\xi}{(\omega t)}}  \,,\ee
where $(\xi)_n=\xi(\xi+1)\cdots(\xi+n-1)$ is the Pochhammer symbol. Normalization of the wave functions gives an identity
\be \cosh^{2\xi}{(\omega t)}=\sum_{n=0}^\infty \fft{(\xi)_n}{n!}\tanh^{2n}(\omega t) \,.\ee
This is particularly useful in the derivation of complexities. We deduce for example
\bea
&& K=\xi \sinh^2(\omega t)\,,\\
&& K_2=\xi^{-1}K\Big( (\xi+1)K+\xi \Big)\,,\nn\\
&& K_3=\xi^{-1}K\Big( (\xi+2)K_2+2(\xi+1)K+\xi \Big)\,,\nn\\
&& K_4=\xi^{-1}K\Big( (\xi+3)K_3+3(\xi+1)K_2+(3\xi+1)K+\xi \Big)\,,\nn\\
&& K_5=\xi^{-1}K\Big( (\xi+4)K_4+(4\xi+6)K_3+(6\xi+4)K_2+(4\xi+1)K+\xi \Big)\,,\nn\\
&& K_6=\xi^{-1}K\Big( (\xi+5)K_5+(5\xi+10)K_4+10(\xi+1)K_3+5(2\xi+1)K_2+(5\xi+1)K+\xi \Big)\,.\nn
\eea
From these results, one finds to leading order in the long time limit
\be K_{\delta+1}=\fft{\xi+\delta}{\xi}\,K K_\delta\quad \Longrightarrow \quad K_\delta=\fft{(\xi)_\delta}{\xi^\delta}\,K^\delta\,.\ee
With this result in hand, we are able to prove the master result (\ref{master}) for SYK-like model explicitly in the subsection \ref{enkk}. Here we shall present several examples. We deduce
\bea
&& \fft{\Delta K_2}{K_2}=\sqrt{\fft{4\xi+6}{\xi(\xi+1)}}\,,\nn\\
&& \fft{\Delta K_3}{K_3}=\sqrt{\fft{9\xi^2+45\xi+60}{\xi(\xi+1)(\xi+2)}}\,.
\eea
Recall $\Delta K/K=1/\sqrt{\xi}$, it follows that the relation (\ref{master}) holds because of a simple fact: $\xi>0$.

\end{document}